\documentclass[showpacs,aps,prl,preprint,onecolumn,superscriptaddress]{revtex4}
\usepackage{graphicx}
\usepackage{dcolumn}
\usepackage{amsmath}
\usepackage{soul,xcolor}
\usepackage[normalem]{ulem}

\begin{document}

\title{Photocurrent imaging of hybrid polaritons in graphene based heterostructures}

\author{Weiwei Luo}
\email{weiwei.luo@nankai.edu.cn}
\affiliation{The Key Laboratory of Weak-Light Nonlinear Photonics, Ministry of Education, School of Physics and TEDA Applied Physics Institute, Nankai University, Tianjin 300457, China}
\affiliation{Collaborative Innovation Center of Extreme Optics, Shanxi University, Taiyuan, Shanxi 030006, China}
\author{Jialin Qi}
\affiliation{The Key Laboratory of Weak-Light Nonlinear Photonics, Ministry of Education, School of Physics and TEDA Applied Physics Institute, Nankai University, Tianjin 300457, China}
\author{Linglong Zhang}
\affiliation{The Key Laboratory of Weak-Light Nonlinear Photonics, Ministry of Education, School of Physics and TEDA Applied Physics Institute, Nankai University, Tianjin 300457, China}
\author{Jiang Fan}
\affiliation{The Key Laboratory of Weak-Light Nonlinear Photonics, Ministry of Education, School of Physics and TEDA Applied Physics Institute, Nankai University, Tianjin 300457, China}
\author{Junjie Dingxiao}
\affiliation{The Key Laboratory of Weak-Light Nonlinear Photonics, Ministry of Education, School of Physics and TEDA Applied Physics Institute, Nankai University, Tianjin 300457, China}
\author{Ni Zhang}
\affiliation{The Key Laboratory of Weak-Light Nonlinear Photonics, Ministry of Education, School of Physics and TEDA Applied Physics Institute, Nankai University, Tianjin 300457, China}
\author{Wei Wu}
\affiliation{The Key Laboratory of Weak-Light Nonlinear Photonics, Ministry of Education, School of Physics and TEDA Applied Physics Institute, Nankai University, Tianjin 300457, China}
\author{Mengxin Ren}
\affiliation{The Key Laboratory of Weak-Light Nonlinear Photonics, Ministry of Education, School of Physics and TEDA Applied Physics Institute, Nankai University, Tianjin 300457, China}
\author{Xinzheng Zhang}
\affiliation{The Key Laboratory of Weak-Light Nonlinear Photonics, Ministry of Education, School of Physics and TEDA Applied Physics Institute, Nankai University, Tianjin 300457, China}
\author{Wei Cai}
\email{weicai@nankai.edu.cn}
\affiliation{The Key Laboratory of Weak-Light Nonlinear Photonics, Ministry of Education, School of Physics and TEDA Applied Physics Institute, Nankai University, Tianjin 300457, China}
\author{Jingjun Xu}
\email{jjxu@nankai.edu.cn}
\affiliation{The Key Laboratory of Weak-Light Nonlinear Photonics, Ministry of Education, School of Physics and TEDA Applied Physics Institute, Nankai University, Tianjin 300457, China}

\date{\today}

\begin{abstract}
Photocurrent is arising as a powerful tool for detecting in-plane collective excitations in hybrid polariton systems. In this paper, based on the intrinsic optoelectric response of graphene, photocurrent imaging of in-plane plasmons from each graphene layer is presented in a hybrid graphene-graphene heterostructure. In combination with near-field optical signals which detect plasmons above the sample, three dimensional detection of hybrid plasmons is demonstrated. Especially, only an electronic boundary is necessary for the electrical detection of hybrid plasmons, which acts as both the photocurrent junction and plasmon reflector. Our studies would promote electrical studies of polariton related physical phenomena and pave the way towards all-electrical nano-optical processing.
\end{abstract}
\maketitle

\setcounter{secnumdepth}{0}
\section{Introduction}
Photocurrent is emerging as a versatile probe of quantum materials with properties governed by physics spanning multiple spatio-temporal scales\cite{ma2022photocurrent}. Graphene plasmons\cite{koppens2011graphene,basov2016polaritons,low2017polaritons}, Dirac electrons coupled with infrared photons, hold ultra confinement and electrical tunability, presenting great opportunities for enhanced light-matter interactions from mid-infrared to teraherz (THz) spectral range. Moreover, probing this collective excitation gives access to the fundamental physical mechanism behind. The appealing optoelectric properties of graphene\cite{gabor2011hot,koppens2014photodetectors} enable intrinsic electrical detection of graphene plasmons\cite{freitag2013photocurrent,torre2015electrical,lundeberg2017thermoelectric,guo2018efficient}. Especially, by combining the broadband nanoscale optical excitation of scattering-type scanning near-field optical microscopy (s-SNOM) with electrical signal readout, nano-photocurrent measurement\cite{woessner2016near} enables electrical imaging of propagating graphene plasmon\cite{lundeberg2017thermoelectric,alonso2017acoustic}, which precludes the difficulties of far-field optical detection, and moreover gets access to novel optoelectric response at nanoscale. Therefore, nonlocal conductivity of graphene\cite{lundeberg2017tuning}, phonon polartions of hBN\cite{woessner2017electrical} and optoelectric response of graphene Moire superlattices\cite{sunku2020nano,hesp2021nano,sunku2021hyperbolic} were demonstrated experimentally. Furthermore, the nonlinear surface conductivity of Weyl metals\cite{shao2021nonlinear} can be revealed from nano-photocurrent imaging.

On the other hand, hybriding graphene plasmon with other polaritons\cite{basov2016polaritons,low2017polaritons,basov2021polariton} provides new freedom of light manipulations and meanwhile an excellent platform for studying intriguing physical phenomena which can be revealed from hybrid polariton dispersions. In hybrid double layer graphene system\cite{hwang2009plasmon}, deeper optical modulations\cite{yan2012tunable,rodrigo2017double} can be realized, in together with the predictions of self-excited plasmons\cite{svintsov2016plasmons,de2017plasmon} and drifting electrons triggered surface plasmon amplifications\cite{morgado2017negative}. Novel plasmon dispersions exist in mixed dimensional graphene heterostructures\cite{jariwala2017mixed,badalyan2017plasmons,hwang2018dimensionally,wang2021gate}. Quantum nonlocal effects of metal\cite{dias2018probing,gonccalves2021quantum} and Higgs modes\cite{costa2021harnessing} can be revealed from hybrid graphene-metal and graphene-superconductor systems, respectively. Nanoscale electrical probe would play an important role in revealing these polariton related phenomena.

In this paper, we present nano-photocurrent studies of hybrid polaritons in a representative graphene-graphene heterostructure. By stacking, an extra vertical dimension is introduced as compared to single layer graphene plasmons. Under vertical electric fields, electronic boundaries (EBs) are created, acting as natural junctions for photocurrent collection. Accordingly, electrical imaging of hybrid polartions from each graphene layer is demonstrated, thus constituting three-dimensional plasmon field imaging in together with near-field optical signals. Moreover, we prove that electrical imaging of polaritons near single EB can be achieved, where the EB acts as both plasmon reflector and photocurrent collector. This achievement simplifies traditional implementations where the reflector and collector are fabricated separately\cite{lundeberg2017thermoelectric,alonso2017acoustic,lundeberg2017tuning,woessner2017electrical}. Our results thus demonstrate flexible approaches for electrical imaging of graphene based hybrid polartions.

\section{Results}
\subsection{Experimental configuration}

\begin{figure*}[htb]
\centerline{\includegraphics[width=12cm]{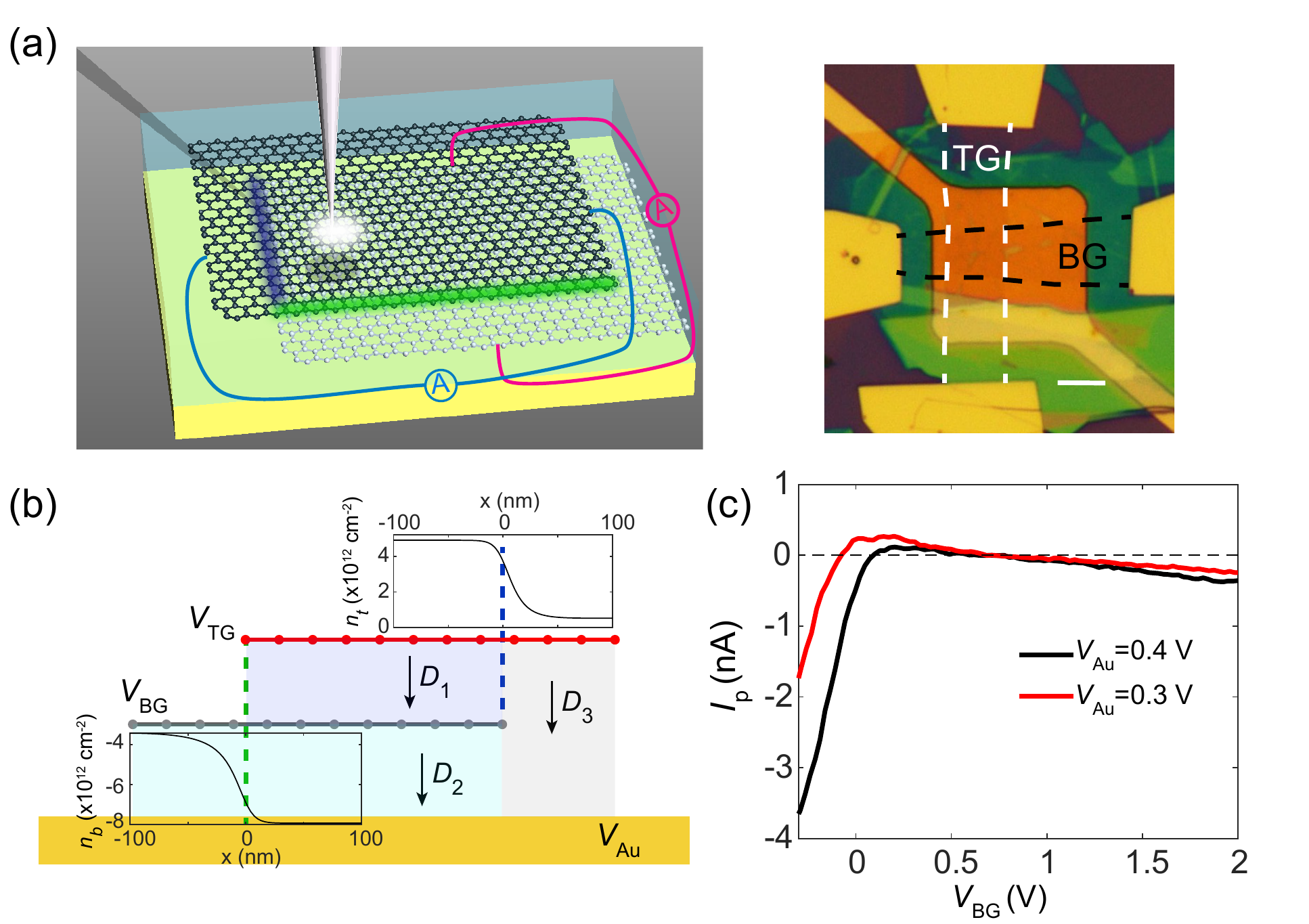}}
\caption{Nano-photocurrent measurements of the hybrid graphene-graphene system. (a) Left: sketch of nano-photocurrent imaging on the hybrid graphene-graphene system. Right: microscope image of the sample. Scar bar, 5~$\mu$m. The system is consisted of two graphene layers encapsulated and separated by thin layers of hBN, sitting on substrate of Au(50~nm)/Cr(5~nm)/silica. Thickness of the top, middle and bottom hBN are 3, 12 and 15~nm, respectively. A mid-infrared laser with wavelength of 10.65~$\mu$m is focused on the metallic tip, and the generated nano-photocurrent signals are collected via the source and drain electrodes on each layer of graphene.  (b) Electrostatic potential induced carrier density distributions on the two graphene layers. $D_1$, $D_2$ and $D_3$ represent the displacement fields existing within TG-BG, BG-Au, and TG-Au, respectively. The top and bottom panels present separately the calculated carrier densities of TG and BG, with $V_{\mathrm{TG}}$=0~V, $V_{\mathrm{BG}}$=3~V and $V_{\mathrm{Au}}$=0.7~V. Electronic boundary is induced on each layer by the vertical projection from edge of the other graphene layer. (c) Sweep of photocurrent  signals at the EB of TG by varying $V_{\mathrm{BG}}$. The two curves are for $V_{\mathrm{Au}}$=0.3 and 0.4~V, respectively.
}
\label{fig1}
\end{figure*}

As illustrated in Fig.\ref{fig1}a, the sample is consisted of two graphene layers which are encapsulated in thin films of hBN with thickness of 3,12 and 15 nm from top to bottom. The whole heterostructure lies on Au. The conducting layer of Au provides back gate and moreover is used for the suppression of photoinduced doping effects from hBN during experiment\cite{ju2014photoinduced,woessner2016near}. In the experiment, a laser wavelength of 10.6~$\mu$m is used, and nano-photocurrent signals $I_\mathrm{p}$ from each layer are collected when the tip scans across the overlapping region, in together with near-field optical signal $s_\mathrm{3}$. $s_\mathrm{3}$ detects electric field from the upper space and can be an excellent reference for comparing $I_\mathrm{p}$ patterns from the two layers.

To detect electrically plasmon response of the hybrid system, junction with different Seebeck coefficients $S$ on the two sides in graphene layer is necessary for the generation of net current $I_\mathrm{p}$. Previously, junction is realized through split back gates which induce doping difference\cite{gabor2011hot,lundeberg2017thermoelectric,lundeberg2017tuning,woessner2017electrical}.
In our experiment, the in-pane dislocation of the two graphene layers naturally introduce EBs under electrical displacement fields. As illustrated in Fig.\ref{fig1}b, electric potentials  $V_{\mathrm{TG}}$ on the top graphene (TG) and $V_{\mathrm{BG}}$ on the bottom graphene (BG) induces displacement field $D_1$ within the overlapping region. On the other hand, back gate voltage on gold $V_{\mathrm{Au}}$ introduce displacement fields $D_2$ under BG and $D_3$ under TG out of the overlapping region. Consequently, EBs of varied carrier density distributions are created on both the TG and BG at the vertical projection of edges of the another layer, naturally inducing different Seeback coefficients on the two sides. In Fig.\ref{fig1}c,  dependence of $I_\mathrm{p}$ across the EB of TG on voltages applied on the two sides is studied. The two curves of nano-photocurrent signals at the EB under two different voltages $V_{\mathrm{Au}}$ both show triple signs with the change of $V_{\mathrm{BG}}$, thus demonstrating a photo-thermoelectric effect (PTE) dominated photocurrent mechanism\cite{gabor2011hot,woessner2016near,lundeberg2017thermoelectric}.

\subsection{Simulation of spatial electrical signal near the EB}
\begin{figure*}[htb]
\centerline{\includegraphics[width=12cm]{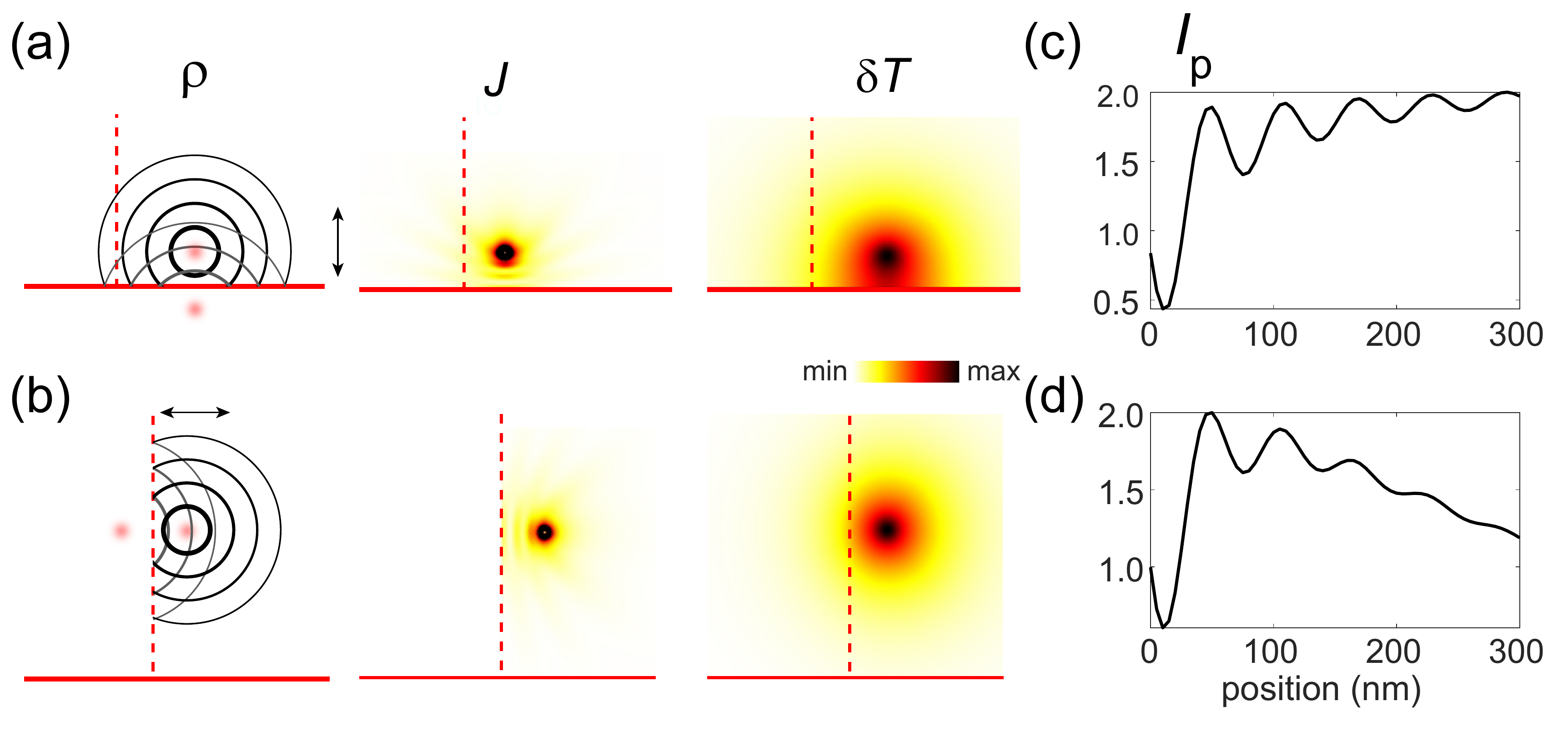}}
\caption{ Simulated electrical probing of plasmon waves. (a,b) Left to right: calculated distributions of carrier $\rho$, absorbed power $J$ and elevated electron temperature $\delta T$ for plasmon reflection near the physical edge (a) and EB (b). For the reflection near the EB (b), plasmon field distribution on the left side of the EB is ignored for simplicity. The red solid lines mark the edge of graphene, and the red dashed lines represent the EB. (c,d) The dependence of $I_\mathrm{p}$ with tip scanning across the physical edge (c) and EB (d). Plasmon reflection coefficients for both cases are both supposed as $r=0.6e^{0.65\pi i}$. Plasmon wavelength is $\lambda_p$=120~nm and damping rate is $\gamma$=0.04. The cooling length $l_\mathrm{c}$ is set as 250~nm.
}
\label{fig2}
\end{figure*}

Besides being junction for collecting $I_\mathrm{p}$, the EBs can also act as plasmon reflectors, as demonstrated from previous near-field optical studies of hybrid graphene-graphene\cite{woessner2017propagating,hu2021direct} and bilayer graphene(BLG)-graphene heterostructures\cite{luo2021nanoinfrared} where plasmon reflections occur because of the impendence mismatch of plasmons across the EBs. Meanwhile, different types of EB have been proved as reflector of graphene plasmons\cite{fei2017nanoplasmonic}, including  grain boundaries\cite{fei2013electronic}, the one induced by ion beam modification\cite{luo2016tailorable,luo2017plane} and carbon nanotube (CNT)\cite{jiang2016tunable}, boundaries between BLG and graphene\cite{alonso2014controlling}, and boundaries between Moire-patterned graphene and graphene\cite{ni2015plasmons}. Here, the efficient reflection of plasmons by the EB can be utilized in electrical imaging of plasmon waves. Fig.\ref{fig2} displays the simulations of $I_\mathrm{p}$ across both the physical edge and EB of graphene, the former being demonstrated previously\cite{lundeberg2017thermoelectric,alonso2017acoustic,lundeberg2017tuning}. For plasmon induced photocurrent in graphene, the in-plane electrical field induce Joule heating, which then diffuses spatially, and increases the electron temperature. After that, photocurrent is generated near the EB, in proportional to the Seebeck coefficient difference $\Delta S$ across the EB and the elevated electron temperature $\delta T$ at the EB. Here, the interference model\cite{lundeberg2017thermoelectric} is introduced for simulating possible patterns of $I_\mathrm{p}$, constructed from a series of equations:
 \begin{eqnarray}
 k_p^{-2}\nabla^2\rho(x,y)+\rho(x,y)=f(x,y),  \\
 J(x,y)\propto|\nabla\rho(x,y)|^2, \\
  \delta T(x,y)-l_c^2\nabla^2\delta T(x,y)=J(x,y)/g.
 \end{eqnarray}
These three equations correspond successively to the excitation of plasmon wave $\rho(x,y)e^{-i\omega t}$, generation of Joule heating $J(x,y)$, and  diffusion of hot carriers. $f(x,y)$ represents the plasmon source of the tip and can be treated as Gaussian distribution of 10~nm width. $k_\mathrm{p}$ is the complex plasmon wave vector, expressed as $k_\mathrm{p}=\frac{2\pi}{\lambda_\mathrm{p}}(1+i\gamma)$, where $\lambda_\mathrm{p}$ is plasmon wavelength, and $\gamma$ features plasmon damping rate. $\delta T$ is the elevated electron temperature, and $l_{\mathrm{c}}=\sqrt{\kappa/g}$ is cooling length characterizing thermal spreading scale, where $\kappa$ and $g$ are the in-plane thermal conductance and out-of-plane heat sinking conductance, respectively.

The left panels in Fig.\ref{fig2}a and b illustrate the excitation and reflection of $\rho$ near the physical edge (red solid lines) and EB (red dashed lines), respectively. In the former case, the reflector (physical edge) and photocurrent collector (EB) are separated. While for the latter one, the EB acts simultaneously these two roles. The calculated distributions of Joule heating $J$ and elevated electron temperature $\delta T$ are presented successively. Afterwards, $I_\mathrm{p}$ under certain tip position can be calculated by integrating $\delta T$ along the EBs.

Figure \ref{fig2}c and d plot the calculated $I_\mathrm{p}$ distributions during scanning tip across the physical edge and EB respectively. Clearly, periodic patterns are observed for both cases, and the periods are both exactly half the plasmon wavelength $\lambda_\mathrm{p}$. Therefore, the reflection of plasmon waves are imaged electrically in both the two cases. The slightly rise of $I_\mathrm{p}$ curve away from the physical edge in Fig.\ref{fig2}c stems from plasmon energy loss near the edge. On the other hand, the competition between the plasmon energy loss and the hot carrier cooling length $l_\mathrm{c}$ results in a relatively flat then rapid decay trend of $I_\mathrm{p}$ near the EB. Additionally, based on the calculated $I_\mathrm{p}$ under different reflection coefficients in Supplementary Fig.3, the oscillation strengths of plasmon waves in both cases are positively correlated to the magnitude of reflection coefficient.

\subsection{Nano-photocurrent experiments near the EB}
\begin{figure*}[htb]
\centerline{\includegraphics[width=13cm]{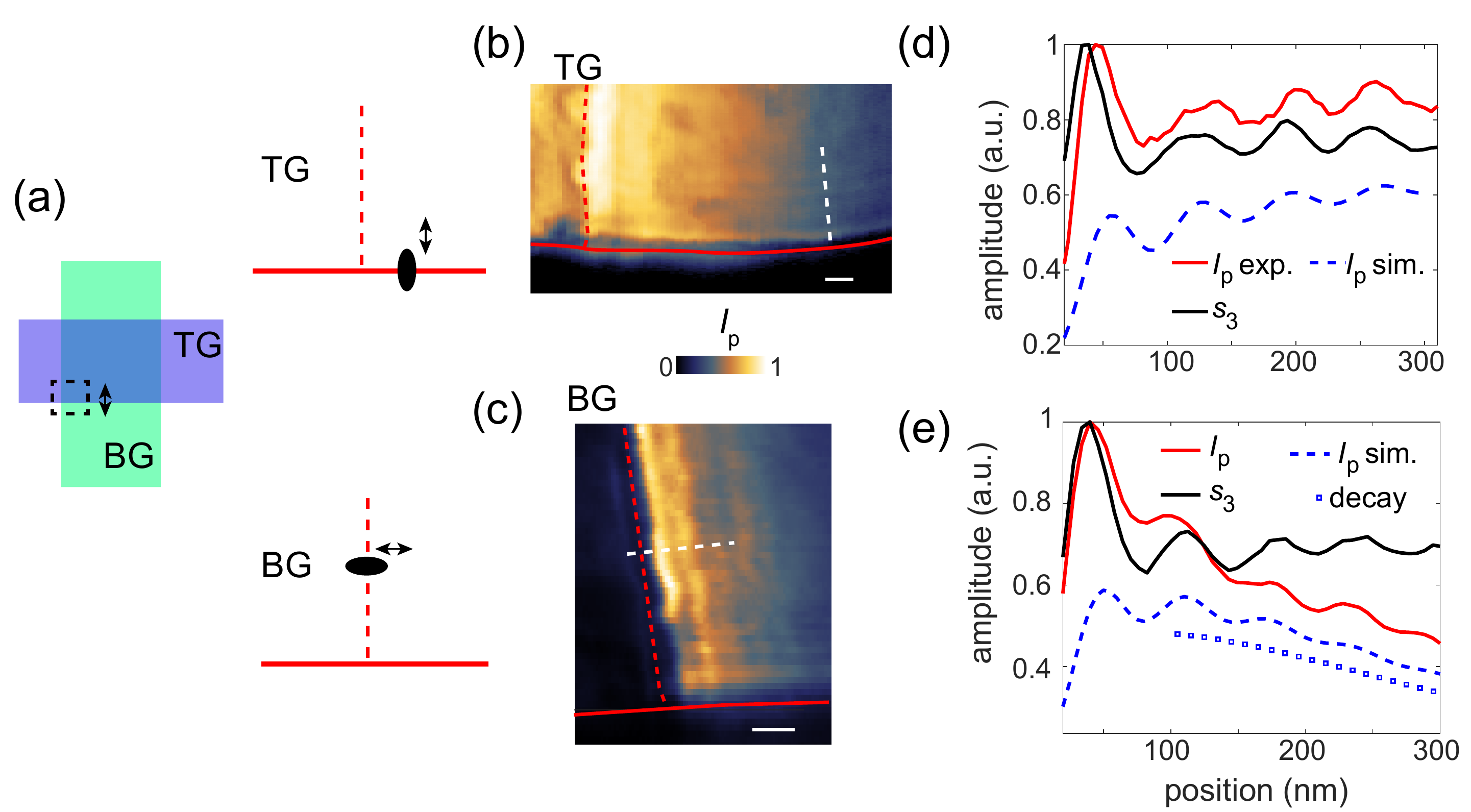}}
\caption{Nano-photocurrent imaging of propagating plasmons. (a) Left: illustration of the scanned region within the graphene-graphene heterostructure. The arrow marks the direction across the edge of TG. Right: the corresponding configurations of electrical detection near the physical edge from the TG (top) and EB of BG (bottom). (b,c) Nano-photocurrent images from the TG (b) and BG (c), respectively. In (a-c), the red solid lines mark the physical edges of the corresponding layer, and the red dashed lines represent the EBs. For (b), $V_{\mathrm{BG}}$=3.0~V, $V_{\mathrm{Au}}$=0.7~V. While for (c), $V_\mathrm{TG}$=3.2~V, $V_{\mathrm{Au}}$=0.7~V. Scale bars, 100~nm. (d,e) Experimentally extracted photocurrent signals $I_\mathrm{p}$ (red solid) and near-field optical signals $s_\mathrm{3}$ (black solid) along the white dashed lines in (b) and (c) respectively, in together with the model fitted $I_\mathrm{p}$ curves (blue dashed). Fitting parameters are $r$=0.4e$^{0.65\pi i}$,$\gamma$=0.04 and $l_\mathrm{c}$=250~nm. The values of $\lambda_\mathrm{p}$ are 140~nm (d) and 126~nm (e), respectively. To calculate the decay curve (blue squre) in (e), $r$ is taken as 0.
}
\label{fig3}
\end{figure*}

Nextly, nano-photocurrent measurements are performed both on the TG and BG (supplementary note 1). For the overlapping region of the heterostructure illustrated in Fig.\ref{fig3}a, the edges of TG and BG induce EBs on the other layer. Corresponding nano-photocurrent results from the TG and BG are presented in Fig.\ref{fig3}b and c, respectively. As expected, $I_\mathrm{p}$ signals from both layers show rapid decay away from each EB. Across the physical edge of TG in Fig.\ref{fig3}b (red solid line), which simultaneously induces EB on BG in Fig.\ref{fig3}c (red dashed line), clear fringes are observed from both graphene layers. The extracted profiles in Fig.\ref{fig3}d and e present consistent spatial patterns between $I_\mathrm{p}$ and $s_3$ for both cases, suggesting detection of reflected plasmons.

\begin{figure*}[htb]
\centerline{\includegraphics[width=14cm]{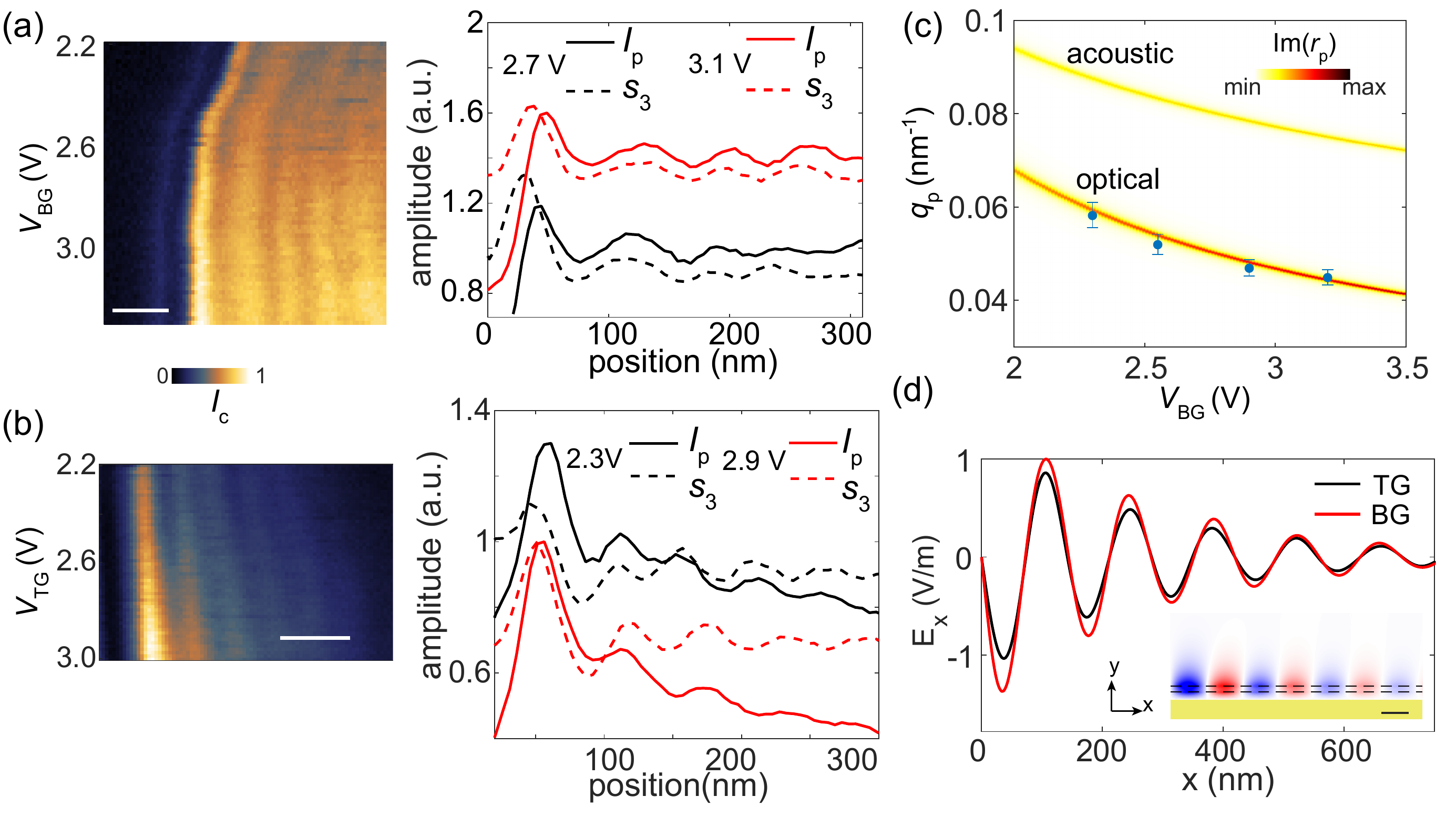}}
\caption{Plasmon dispersions from nano-photocurrent imaging. (a) Left: dependence of $I_\mathrm{p}$ across the edge from TG under $V_{\mathrm{BG}}$=3.3 to 2.2~V and $V_{\mathrm{Au}}$=0.7~V. Right: extracted profiles of $I_\mathrm{p}$ and $s_3$ at $V_{\mathrm{BG}}$=3.1 and 2.7~V. Scar bar, 100~nm. (b) The case for $I_\mathrm{p}$ across the EB from BG under $V_{\mathrm{TG}}$=3.0 to 2.2~V and $V_{\mathrm{Au}}$=-0.4~V.  Scar bar, 100~nm. The curves in (a,b) are offset for comparison. (c) Plasmon dispersion calculated from the imaginary part of reflection coefficient $r_p$ for the case in (a). The blue dots with error bars are the experimental results obtained from $I_\mathrm{p}$. (d) Simulated distribution of electric field $E_\mathrm{x}$  within the two graphene layers. Inset: distributions of $E_\mathrm{x}$ within the heterostructure. The two dashed lines mark the two graphene layers. Conductivities of the TG and BG are taken as (0.15+2.6i)$G_0$ and (0.15+3.3i)$G_0$, respectively, where $G_0=q_e^2/4\hbar$. A vertical line dipole 100~nm above is utilized as the source. Scale bar, 50~nm.
}
\label{fig4}
\end{figure*}

To prove the electrical imaging of propagating plasmons, voltage dependent nano-photocurrent signals are measured and presented in Fig.\ref{fig4} a and b, respectively. Clear dispersions of $I_\mathrm{p}$ are observed in both cases, in consistent with near-field signals $s_3$. Accordingly, plasmon wavelength can be extracted as twice the period of the fringes. Fig.\ref{fig4}c presents the calculated dispersion of the hybrid plasmons, which is consisted of the branches of optical and acoustic modes\cite{hwang2009plasmon}. As shown, the experimental data fits the optical mode quite well. Therefore, the propagating optical plasmon mode is electrically imaged from both the TG and BG in Fig.\ref{fig3}. In combination with the near-field optical signals $s_3$ which detect electrical field above the sample, a three dimensional electrical imaging of hybrid plasmons covering both in-plane and vertical dimensions are demonstrated here. The observed consistent plasmon patterns among $I_\mathrm{p}$ from both layers and $s_3$ can be followed by the simulated field distributions in Fig.\ref{fig4}d. For a dipole source located above, the two graphene layers possess nearly consistent in-plane plasmon oscillations, which are mostly contributed from the hybrid optical mode. In addition, previous s-SNOM experiments have shown that $s_3$ detects mostly the hybrid optical mode\cite{woessner2017propagating,hu2021direct,luo2021nanoinfrared}. The acoustic mode can be barely distinguished in experiment, being related to the higher excitation efficiency for lower in-plane wave momentum from the metallic tip. Besides, plasmon fields of the acoustic mode are confined mostly within the gap of the two graphene layers, and couple to the tip weakly as compared to the optical mode where electrical fields relies mostly outside of the gap between graphene layers.

Meanwhile, the detected plasmon fringes are near the edge (Fig.\ref{fig3}d) and EB (Fig.\ref{fig3}e) of graphene layers respectively, verifying the corresponding simulation results in Fig.\ref{fig2}. The reflection of plasmon waves stems from the distinct plasmon dispersions on the two sides. The experimental $I_\mathrm{p}$ curves can be fitted, as shown in Fig.\ref{fig3}d and e. For both cases, the spatial patterns of plasmon waves are matched quite well, with a plasmon damping rate $\gamma$=0.04, which is expected for hBN encapsulated graphene\cite{woessner2017propagating}. Moreover, the decay trend of $I_\mathrm{p}$ away from the EB (Fig.\ref{fig3}e) is captured well in the model (blue squre curve), yielding a fitted cooling length of 250~nm, a value similar to the previous study\cite{lundeberg2017thermoelectric}. Divergence between experiment and simulation near the first peaks can be attributed to the contribution from the other side of the EB, which is not considered in the model.

On the other hand, near the another boundary of the overlapping region in Fig.\ref{fig3}, which is the edge of BG (red dashed curve in Fig.\ref{fig3}b and red solid line in Fig.\ref{fig3}c), the plasmon interference patterns are weaker, indicating lower reflection coefficient. The varied reflection coefficients near the two boundaries are related to the quality of the physical edges during transferring and the impendence difference across the boundaries. The reflection coefficient can be increased by reducing gap between graphene layers and improving sample quality.

\section{Discussion}
In conclusion, electrical imaging of hybrid plamons in graphene-graphene heterostructures is demonstrated from both nano-photocurrent measurements on the top and bottom graphene layers. In combination with scattering near-field optical signals, three dimensional imaging of plasmon waves are presented for the first time. Moreover, it is proved that photocurrent imaging of propagating plasmons can be achieved near the single EB which act simultaneously as photocurrent junction and plasmon reflector. These results provide significant approaches for flexible electrical probe of hybrid polaritons in graphene based heterostructures, where in-plane electrical field distributions in different planes can be detected by adjusting the location of graphene layer. Meanwhile, unloosing the traditional restrictions where the junction and reflector are individually fabricated would benefit greatly the design of device architecture. Electronic boundaries with efficient reflection coefficient have been demonstrated in different configurations\cite{fei2017nanoplasmonic}, and can be induced in systems, including graphene with split gate, mixed-dimensional graphene heterostructures\cite{jiang2016tunable,wang2021gate}, graphene-graphene\cite{woessner2017propagating,li2020global,hu2021direct,luo2021nanoinfrared}, graphene-MoS$_2$\cite{sunku2021dual}, graphene-WSe$_2$\cite{hesp2022wse2}, graphene-superconductor\cite{costa2021harnessing} and other heterostructures based on them.The demonstrated flexible electrical imaging of hybrid polartions would pave the way towards all-electrical nano-optical processing and promote the studies of novel physical phenomena.

\section{Materials and methods}
\subsection{Sample preparation}
Graphene and h-BN are obtained by mechanical cleavage of bulk graphite and hexagonal boron nitride crystal.
Before the transferring, back gate of Au(50~nm)/Cr(7~nm) are deposited on SiO$_2$ (285~nm)/Si by UV-lithography and electron-beam evaporation. During the sample transferring, stamps of PDMS (polydimethysiloxane)/polycarbonate (PC) are used for the picking up and stacking of different layers of the heterostructure. Assembly of  the stack is divided into three steps: transferring bottom hBN, middle hBN/bottom layer of graphene, and top hBN/top layer of graphene successively. For each step, the layers attached on the stamps are released on the desired locations at a temperature around 150$^{\circ}$, followed by dissolution of PC film in chloroform. Finally,  One dimensional electrical contacts of Au(70~nm)/Cr(7~nm) to the two graphene layers are realized by electron beam lithography, reactive ion etching (mixed gases of CHF$_3$ and O$_2$) and electron-beam deposition.

\subsection{Nano-photocurrent configuration and signal analysis}
Nano-photocurrent experiment is conducted based on scattering-type scanning near-field microscopy (s-SNOM, neaspec), equipped with a CO$_2$ laser (wavelength of 9.3-10.7~$\mu$m). The oscillating metallic tip with frequency of around 250~kHz excites photocurrent signals $I_\mathrm{p}$ on the sample, which are collected and amplified with a current amplifier (FEMTO DHPCA-100) and demodulated at different harmonics $n$ of the tip frequency. In our experiment, $n=1$ and $n=2$ are employed (Supplementary note 2).

During the experiment, both the amplitude and phase channels of photocurrent signals are obtained. $I_\mathrm{p}$ is obtained as the real part of the complex signals. However, before that, the phase channel needs to be corrected\cite{hesp2021nano}. For different harmonics $n$, $n$ times of the mechanical phase signals are subtracted from the phase channel. Then the most frequent phase within a scan is deducted, which is due to the electronic decay in a circuit.

\subsection{Data availability}
The data that support the findings of this study are available from the corresponding author upon reasonable request.

\section{Acknowledgements}
W. L. thanks Chunyan Jin for advices on electron beam lithography. Funding: This work has been supported by Guangdong Major Project of
Basic and Applied Basic Research (2020B0301030009), the National Natural Science Foundation of China (12004196, 12074200, 12127803, 12074201,12222408), Changjiang Scholars and Innovative Research Team in University (IRT13\_R29), the 111 Project (B07013), and Fundamental
Research Funds for the Central Universities.

\section{Conflict of interest}
The authors declare no competing interests.

\section{Contributions}
W.L.,W.C., and J.X. conceived the idea and supervised the project. W.L., J.Q.and J.D. prepared the sample. W.L., J.Q. and L. Z. performed
the experiments. L.W. conducted the theoretical simulations.  All authors discussed the data and contributed to writing the manuscript.


\begin{thebibliography}{10}
\expandafter\ifx\csname url\endcsname\relax
  \def\url#1{\texttt{#1}}\fi
\expandafter\ifx\csname urlprefix\endcsname\relax\def\urlprefix{URL }\fi
\expandafter\ifx\csname href\endcsname\relax
  \def\href#1#2{#2} \def\path#1{#1}\fi

\bibitem{ma2022photocurrent}
Q.~Ma, R.~K. Kumar, S.-Y. Xu, F.~H. Koppens, J.~C. Song, Photocurrent as a
  multi-physics diagnostic of quantum materials, arXiv preprint
  arXiv:2210.13485 (2022).

\bibitem{koppens2011graphene}
F.~H. Koppens, D.~E. Chang, F.~J. Garc{\'\i}a~de Abajo, Graphene plasmonics: a
  platform for strong light--matter interactions, Nano Letters 11~(8) (2011)
  3370--3377.

\bibitem{basov2016polaritons}
D.~Basov, M.~Fogler, F.~Garc{\'\i}a~de Abajo, Polaritons in van der waals
  materials, Science 354~(6309) (2016) aag1992.

\bibitem{low2017polaritons}
T.~Low, A.~Chaves, J.~D. Caldwell, A.~Kumar, N.~X. Fang, P.~Avouris, T.~F.
  Heinz, F.~Guinea, L.~Martin-Moreno, F.~Koppens, Polaritons in layered
  two-dimensional materials, Nature Materials 16~(2) (2017) 182--194.

\bibitem{gabor2011hot}
N.~M. Gabor, J.~C. Song, Q.~Ma, N.~L. Nair, T.~Taychatanapat, K.~Watanabe,
  T.~Taniguchi, L.~S. Levitov, P.~Jarillo-Herrero, Hot carrier--assisted
  intrinsic photoresponse in graphene, Science 334~(6056) (2011) 648--652.

\bibitem{koppens2014photodetectors}
F.~Koppens, T.~Mueller, P.~Avouris, A.~Ferrari, M.~Vitiello, M.~Polini,
  Photodetectors based on graphene, other two-dimensional materials and hybrid
  systems, Nature Nanotechnology 9~(10) (2014) 780--793.

\bibitem{freitag2013photocurrent}
M.~Freitag, T.~Low, W.~Zhu, H.~Yan, F.~Xia, P.~Avouris, Photocurrent in
  graphene harnessed by tunable intrinsic plasmons, Nature Communications 4~(1)
  (2013) 1--8.

\bibitem{torre2015electrical}
I.~Torre, A.~Tomadin, R.~Krahne, V.~Pellegrini, M.~Polini, Electrical plasmon
  detection in graphene waveguides, Physical Review B 91~(8) (2015) 081402.

\bibitem{lundeberg2017thermoelectric}
M.~B. Lundeberg, Y.~Gao, A.~Woessner, C.~Tan, P.~Alonso-Gonz{\'a}lez,
  K.~Watanabe, T.~Taniguchi, J.~Hone, R.~Hillenbrand, F.~H. Koppens,
  Thermoelectric detection and imaging of propagating graphene plasmons, Nature
  Materials 16~(2) (2017) 204--207.

\bibitem{guo2018efficient}
Q.~Guo, R.~Yu, C.~Li, S.~Yuan, B.~Deng, F.~J. Garc{\'\i}a~de Abajo, F.~Xia,
  Efficient electrical detection of mid-infrared graphene plasmons at room
  temperature, Nature Materials 17~(11) (2018) 986--992.

\bibitem{woessner2016near}
A.~Woessner, P.~Alonso-Gonz{\'a}lez, M.~B. Lundeberg, Y.~Gao, J.~E.
  Barrios-Vargas, G.~Navickaite, Q.~Ma, D.~Janner, K.~Watanabe, A.~W. Cummings,
  et~al., Near-field photocurrent nanoscopy on bare and encapsulated graphene,
  Nature Communications 7~(1) (2016) 1--7.

\bibitem{alonso2017acoustic}
P.~Alonso-Gonz{\'a}lez, A.~Y. Nikitin, Y.~Gao, A.~Woessner, M.~B. Lundeberg,
  A.~Principi, N.~Forcellini, W.~Yan, S.~V{\'e}lez, A.~Huber, et~al., Acoustic
  terahertz graphene plasmons revealed by photocurrent nanoscopy, Nature
  Nanotechnology 12~(1) (2017) 31--35.

\bibitem{lundeberg2017tuning}
M.~B. Lundeberg, Y.~Gao, R.~Asgari, C.~Tan, B.~Van~Duppen, M.~Autore,
  P.~Alonso-Gonz{\'a}lez, A.~Woessner, K.~Watanabe, T.~Taniguchi, et~al.,
  Tuning quantum nonlocal effects in graphene plasmonics, Science 357~(6347)
  (2017) 187--191.

\bibitem{woessner2017electrical}
A.~Woessner, R.~Parret, D.~Davydovskaya, Y.~Gao, J.-S. Wu, M.~B. Lundeberg,
  S.~Nanot, P.~Alonso-Gonz{\'a}lez, K.~Watanabe, T.~Taniguchi, et~al.,
  Electrical detection of hyperbolic phonon-polaritons in heterostructures of
  graphene and boron nitride, npj 2D Materials and Applications 1~(1) (2017)
  1--6.

\bibitem{sunku2020nano}
S.~S. Sunku, A.~S. McLeod, T.~Stauber, H.~Yoo, D.~Halbertal, G.~Ni,
  A.~Sternbach, B.-Y. Jiang, T.~Taniguchi, K.~Watanabe, et~al.,
  Nano-photocurrent mapping of local electronic structure in twisted bilayer
  graphene, Nano letters 20~(5) (2020) 2958--2964.

\bibitem{hesp2021nano}
N.~C. Hesp, I.~Torre, D.~Barcons-Ruiz, H.~Herzig~Sheinfux, K.~Watanabe,
  T.~Taniguchi, R.~Krishna~Kumar, F.~H. Koppens, Nano-imaging photoresponse in
  a moir{\'e} unit cell of minimally twisted bilayer graphene, Nature
  Communications 12~(1) (2021) 1--8.

\bibitem{sunku2021hyperbolic}
S.~S. Sunku, D.~Halbertal, T.~Stauber, S.~Chen, A.~S. McLeod, A.~Rikhter, M.~E.
  Berkowitz, C.~F.~B. Lo, D.~E. Gonzalez-Acevedo, J.~C. Hone, et~al.,
  Hyperbolic enhancement of photocurrent patterns in minimally twisted bilayer
  graphene, Nature Communications 12~(1) (2021) 1--7.

\bibitem{shao2021nonlinear}
Y.~Shao, R.~Jing, S.~H. Chae, C.~Wang, Z.~Sun, E.~Emmanouilidou, S.~Xu,
  D.~Halbertal, B.~Li, A.~Rajendran, et~al., Nonlinear nanoelectrodynamics of a
  {W}eyl metal, Proceedings of the National Academy of Sciences 118~(48) (2021)
  e2116366118.

\bibitem{basov2021polariton}
D.~Basov, A.~Asenjo-Garcia, P.~J. Schuck, X.~Zhu, A.~Rubio, Polariton panorama,
  Nanophotonics 10~(1) (2021) 549--577.

\bibitem{hwang2009plasmon}
E.~Hwang, S.~D. Sarma, Plasmon modes of spatially separated double-layer
  graphene, Physical Review B 80~(20) (2009) 205405.

\bibitem{yan2012tunable}
H.~Yan, X.~Li, B.~Chandra, G.~Tulevski, Y.~Wu, M.~Freitag, W.~Zhu, P.~Avouris,
  F.~Xia, Tunable infrared plasmonic devices using graphene/insulator stacks,
  Nature Nanotechnology 7~(5) (2012) 330--334.

\bibitem{rodrigo2017double}
D.~Rodrigo, A.~Tittl, O.~Limaj, F.~Abajo, V.~Pruneri, H.~Altug, Double-layer
  graphene for enhanced tunable infrared plasmonics, Light: Science \&
  Applications 6~(6) (2017) e16277--e16277.

\bibitem{svintsov2016plasmons}
D.~Svintsov, Z.~Devizorova, T.~Otsuji, V.~Ryzhii, Plasmons in tunnel-coupled
  graphene layers: Backward waves with quantum cascade gain, Physical Review B
  94~(11) (2016) 115301.

\bibitem{de2017plasmon}
S.~de~Vega, F.~J. Garcia~de Abajo, Plasmon generation through electron
  tunneling in graphene, ACS Photonics 4~(9) (2017) 2367--2375.

\bibitem{morgado2017negative}
T.~A. Morgado, M.~G. Silveirinha, Negative landau damping in bilayer graphene,
  Physical Review Letters 119~(13) (2017) 133901.

\bibitem{jariwala2017mixed}
D.~Jariwala, T.~J. Marks, M.~C. Hersam, Mixed-dimensional van der waals
  heterostructures, Nature Materials 16~(2) (2017) 170--181.

\bibitem{badalyan2017plasmons}
S.~Badalyan, A.~Shylau, A.-P. Jauho, Plasmons in dimensionally mismatched
  coulomb coupled graphene systems, Physical Review Letters 119~(12) (2017)
  126801.

\bibitem{hwang2018dimensionally}
E.~Hwang, B.~Y.-K. Hu, S.~D. Sarma, Dimensionally mixed coupled collective
  modes, Physical Review B 98~(16) (2018) 161304.

\bibitem{wang2021gate}
S.~Wang, S.~Yoo, S.~Zhao, W.~Zhao, S.~Kahn, D.~Cui, F.~Wu, L.~Jiang, M.~Utama,
  H.~Li, et~al., Gate-tunable plasmons in mixed-dimensional van der waals
  heterostructures, Nature Communications 12~(1) (2021) 1--7.

\bibitem{dias2018probing}
E.~J. Dias, D.~A. Iranzo, P.~Gon{\c{c}}alves, Y.~Hajati, Y.~V. Bludov, A.-P.
  Jauho, N.~A. Mortensen, F.~H. Koppens, N.~Peres, Probing nonlocal effects in
  metals with graphene plasmons, Physical Review B 97~(24) (2018) 245405.

\bibitem{gonccalves2021quantum}
P.~Gon{\c{c}}alves, T.~Christensen, N.~M. Peres, A.-P. Jauho, I.~Epstein, F.~H.
  Koppens, M.~Solja{\v{c}}i{\'c}, N.~A. Mortensen, Quantum surface-response of
  metals revealed by acoustic graphene plasmons, Nature Communications 12~(1)
  (2021) 1--7.

\bibitem{costa2021harnessing}
A.~Costa, P.~Gon{\c{c}}alves, D.~Basov, F.~H. Koppens, N.~A. Mortensen,
  N.~Peres, Harnessing ultraconfined graphene plasmons to probe the
  electrodynamics of superconductors, Proceedings of the National Academy of
  Sciences 118~(4) (2021) e2012847118.

\bibitem{ju2014photoinduced}
L.~Ju, J.~Velasco, E.~Huang, S.~Kahn, C.~Nosiglia, H.-Z. Tsai, W.~Yang,
  T.~Taniguchi, K.~Watanabe, Y.~Zhang, et~al., Photoinduced doping in
  heterostructures of graphene and boron nitride, Nature nanotechnology 9~(5)
  (2014) 348--352.

\bibitem{woessner2017propagating}
A.~Woessner, A.~Misra, Y.~Cao, I.~Torre, A.~Mishchenko, M.~B. Lundeberg,
  K.~Watanabe, T.~Taniguchi, M.~Polini, K.~S. Novoselov, et~al., Propagating
  plasmons in a charge-neutral quantum tunneling transistor, ACS Photonics
  4~(12) (2017) 3012--3017.

\bibitem{hu2021direct}
C.~Hu, A.~Deng, P.~Shen, X.~Luo, X.~Zhou, T.~Wu, X.~Huang, Y.~Dong,
  K.~Watanabe, T.~Taniguchi, et~al., Direct imaging of interlayer-coupled
  symmetric and antisymmetric plasmon modes in graphene/hbn/graphene
  heterostructures, Nanoscale 13~(35) (2021) 14628--14635.

\bibitem{luo2021nanoinfrared}
W.~Luo, A.~B. Kuzmenko, J.~Qi, N.~Zhang, W.~Wu, M.~Ren, X.~Zhang, W.~Cai,
  J.~Xu, Nanoinfrared characterization of bilayer graphene conductivity under
  dual-gate tuning, Nano Letters 21~(12) (2021) 5151--5157.

\bibitem{fei2017nanoplasmonic}
Z.~Fei, G.-X. Ni, B.-Y. Jiang, M.~M. Fogler, D.~Basov, Nanoplasmonic phenomena
  at electronic boundaries in graphene, ACS Photonics 4~(12) (2017) 2971--2977.

\bibitem{fei2013electronic}
Z.~Fei, A.~Rodin, W.~Gannett, S.~Dai, W.~Regan, M.~Wagner, M.~Liu, A.~McLeod,
  G.~Dominguez, M.~Thiemens, et~al., Electronic and plasmonic phenomena at
  graphene grain boundaries, Nature Nanotechnology 8~(11) (2013) 821--825.

\bibitem{luo2016tailorable}
W.~Luo, W.~Cai, W.~Wu, Y.~Xiang, M.~Ren, X.~Zhang, J.~Xu, Tailorable reflection
  of surface plasmons in defect engineered graphene, 2D Materials 3~(4) (2016)
  045001.

\bibitem{luo2017plane}
W.~Luo, W.~Cai, Y.~Xiang, W.~Wu, B.~Shi, X.~Jiang, N.~Zhang, M.~Ren, X.~Zhang,
  J.~Xu, In-plane electrical connectivity and near-field concentration of
  isolated graphene resonators realized by ion beams, Advanced Materials
  29~(30) (2017) 1701083.

\bibitem{jiang2016tunable}
B.-Y. Jiang, G.~Ni, C.~Pan, Z.~Fei, B.~Cheng, C.~N. Lau, M.~Bockrath, D.~N.
  Basov, M.~M. Fogler, Tunable plasmonic reflection by bound 1{D} electron
  states in a 2{D} {D}irac metal, Physical Review Letters 117~(8) (2016)
  086801.

\bibitem{alonso2014controlling}
P.~Alonso-Gonz{\'a}lez, A.~Y. Nikitin, F.~Golmar, A.~Centeno, A.~Pesquera,
  S.~V{\'e}lez, J.~Chen, G.~Navickaite, F.~Koppens, A.~Zurutuza, et~al.,
  Controlling graphene plasmons with resonant metal antennas and spatial
  conductivity patterns, Science 344~(6190) (2014) 1369--1373.

\bibitem{ni2015plasmons}
G.~Ni, H.~Wang, J.~Wu, Z.~Fei, M.~Goldflam, F.~Keilmann, B.~{\"O}zyilmaz,
  A.~Castro~Neto, X.~Xie, M.~Fogler, et~al., Plasmons in graphene moir{\'e}
  superlattices, Nature Materials 14~(12) (2015) 1217--1222.

\bibitem{li2020global}
H.~Li, M.~I.~B. Utama, S.~Wang, W.~Zhao, S.~Zhao, X.~Xiao, Y.~Jiang, L.~Jiang,
  T.~Taniguchi, K.~Watanabe, et~al., Global control of stacking-order phase
  transition by doping and electric field in few-layer graphene, Nano Letters
  20~(5) (2020) 3106--3112.

\bibitem{sunku2021dual}
S.~S. Sunku, D.~Halbertal, R.~Engelke, H.~Yoo, N.~R. Finney, N.~Curreli, G.~Ni,
  C.~Tan, A.~S. McLeod, C.~F.~B. Lo, et~al., Dual-gated graphene devices for
  near-field nano-imaging, Nano Letters 21~(4) (2021) 1688--1693.

\bibitem{hesp2022wse2}
N.~C. Hesp, M.~K. Svendsen, K.~Watanabe, T.~Taniguchi, K.~S. Thygesen,
  I.~Torre, F.~H. Koppens, W{S}e$_2$ as transparent top gate for infrared
  near-field microscopy, Nano Letters 22~(15) (2022) 6200--6206.

\end{thebibliography}
\end{document}